\newcommand{\rem}[1]{}
\newcommand{\beq}{\begin{equation}}
\newcommand{\eeq}{\end{equation}}
\newcommand{\beqa}{\begin{eqnarray}}
\newcommand{\eeqa}{\end{eqnarray}}
\newcommand{\ba}{\begin{array}}
\newcommand{\ea}{\end{array}}

\documentclass[pra,twocolumn,showpacs,preprintnumbers,
amsmath,amssymb,floatfix]{revtex4}
\usepackage{hyperref}
\usepackage{epsfig}

\begin{document}

\title{Condensate Fraction of a Two-Dimensional Attractive Fermi Gas} 

\author{Luca Salasnich}
\affiliation{CNR-INFM and CNISM, 
Dipartimento di Fisica ``Galileo Galilei'', 
Universit\`a di Padova, Via Marzolo 8, 35131 Padova, Italy}

\date{\today}

\begin{abstract}
We investigate the Bose-Einstein condensation of fermionic pairs in 
a two-dimensional uniform two-component Fermi superfluid 
obtaining an explicit formula 
for the condensate density as a function of the chemical potential 
and the energy gap. By using the mean-field extended BCS  
theory, we analyze, as a function of the bound-state 
energy, the off-diagonal long-range order in the crossover 
from the Bardeen-Cooper-Schrieffer (BCS) state 
of weakly-bound Cooper pairs to the Bose-Einstein Condensate (BEC) 
of strongly-bound molecular dimers. 
\end{abstract}


\maketitle

In the last few years several 
experimental groups have observed 
the crossover from the Bardeen-Cooper-Schrieffer (BCS) 
state of Cooper Fermi pairs 
to the Bose-Einstein condensate (BEC) 
of molecular dimers with ultra-cold two hyperfine component Fermi 
vapors of $^{40}$K atoms \cite{greiner,regal,kinast} 
and $^6$Li atoms \cite{zwierlein,chin}. 
A finite condensed fraction has been detected 
also in the BCS side 
of the crossover \cite{regal,zwierlein}, 
stimulating a debate over its interpretation
\cite{falco,avdeenkov,diener,pps}. 
Extended BCS (EBCS) equations \cite{eagles,leggett,noziers} 
have been used to reproduce density profiles \cite{perali} and collective 
oscillations \cite{hu} of these Fermi gases. 
Improvements based on Monte Carlo fitting and superfluid 
dynamics have been considered \cite{manini05,others}, 
showing that EBCS is quite accurate. 
As this EBCS mean-field theory is defined for any value of the coupling, it
provides an interpolation between the BCS weak-coupling regime and the BEC
strong-coupling limit \cite{sademelo,engelbrecht}. 
Despite well know limitations \cite{sademelo} the EBCS theory 
is considered a reliable approximation for studying the whole 
BCS-BEC crossover at zero temperature, giving a simple and 
coherent description of the crossover in terms of fermionic variables. 
\par 
Recently, within the EBCS scheme, we have derived \cite{sala-odlro} 
an explicit formula for the number of 
condensed fermionic pairs in the uniform BCS ground-state. 
We have used the EBCS equations to study 
the behavior of this condensate fraction 
as a function of the inter-atomic scattering length in the BCS-BEC
crossover: from the BCS regime crossing the 
unitarity limit to the BEC regime. 
With no fitted parameters, we have found a remarkable 
agreement with experimental results \cite{zwierlein} 
indicating a relevant fraction of condensed pairs of $^6$Li 
atoms also on the BCS side of the Feshbach resonance. 
Note that our analytic formula \cite{sala-odlro} 
of the condensed fraction of the three-dimensional (3D) 
attractive Fermi gas has been obtained independently 
also by Ortiz and Dukelsky \cite{ortiz} and Ohashi and 
Griffin \cite{ohashi}. The formula has been then compared with 
Monte Carlo calculations by 
Astrakharchik {\it et al.} \cite{astrakharchik}: 
these Monte Carlo results show that the analytical formula slightly 
overestimates the condensed fraction of Fermi pairs. 
\par 
For repulsive Fermi gases, it has been predicted that a reduced 
dimensionality strongly modifies density profiles 
\cite{schneider,sala0,sala1,vignolo}, 
collective modes \cite{minguzzi} and stability of mixtures 
\cite{das,sala2}. In this paper we calculate 
the condensate fraction of Fermi pairs in a strictly 2D attractive 
Fermi system. It is well known that 
purely attractive potentials have bound states 
in 1D and 2D for any strength, contrary to the 3D case \cite{landau}. 
It follows that a bound state appears 
immediately as the two-body attraction 
is introduced in a 2D Fermi gas \cite{altri,nussinov}. 
As discussed by Marini, Pistolesi and Strinati \cite{marini}, 
for a 2D Fermi superfluid in the EBCS theory it is 
the value of the bound-state energy that determines the crossover from 
BCS state of weakly-bound Cooper pairs to the BEC 
of strongly-bound dimers. We shall show that the condensate fraction 
of Fermi pairs, which can be expressed in terms of 
elementary functions, increases smoothly during 
the 2D BCS-BEC crossover and becomes equal to $1/2$ 
(all pairs are condensed) only for a very large 
bound-state energy. 

The Hamiltonian density of a dilute and 
interacting two-spin-component Fermi 
gas in a box of volume $V$ is given by
\beq 
{\hat {\cal H}} = -{\hbar^2\over 2 m}
\sum_{\sigma=\uparrow, \downarrow} 
{\hat \psi}^+_{\sigma} \nabla^2 {\hat \psi}_{\sigma} 
+ g \; {\hat \psi}^+_{\uparrow}
{\hat \psi}^+_{\downarrow}
{\hat \psi}_{\downarrow}
{\hat \psi}_{\uparrow} \; , 
\label{ham} 
\eeq 
where ${\hat \psi}_{\sigma}({\bf r})$ is the field operator 
that destroys a Fermion of spin $\sigma$ 
in the position ${\bf r}$, while ${\hat \psi}_{\sigma}^+({\bf r})$ 
creates a Fermion of spin $\sigma$ in ${\bf r}$. 
The attractive inter-atomic interaction is described by a contact 
pseudo-potential of strength $g$ ($g<0$). 
The number density operator is 
\beq 
{\hat n}({\bf r}) = \sum_{\sigma=\uparrow, \downarrow}
{\hat \psi}^+_{\sigma}({\bf r}){\hat \psi}_{\sigma}({\bf r}) 
\eeq  
and the average number of fermions reads 
\beq 
N=\int d^3{\bf r}\, \langle {\hat n}({\bf r})\rangle \; . 
\eeq

The interacting term can be treated within the mean-field 
Hartree-Fock approximation, namely  
\beqa
{\hat \psi}^+_{\uparrow}
{\hat \psi}^+_{\downarrow}
{\hat \psi}_{\downarrow}
{\hat \psi}_{\uparrow} 
= 
\langle 
{\hat \psi}^+_{\uparrow}{\hat \psi}^+_{\downarrow}
\rangle 
{\hat \psi}_{\downarrow}
{\hat \psi}_{\uparrow} 
+ 
{\hat \psi}^+_{\uparrow}
{\hat \psi}^+_{\downarrow} 
\langle 
{\hat \psi}_{\downarrow}{\hat \psi}_{\uparrow} 
\rangle 
\\ \nonumber
+ 
\langle 
{\hat \psi}^+_{\uparrow}
{\hat \psi}_{\uparrow}
\rangle 
{\hat \psi}^+_{\downarrow}
{\hat \psi}_{\downarrow}
+ 
{\hat \psi}^+_{\uparrow}
{\hat \psi}_{\uparrow}
\langle
{\hat \psi}^+_{\downarrow}
{\hat \psi}_{\downarrow}
\rangle 
\eeqa 
and the Hamiltonian density (\ref{ham}) 
is diagonalized by using the following Bogoliubov 
representation of the field operator 
\beq 
{\hat \psi}_{\uparrow}({\bf r})= {1\over V^{1/2}} \sum_{\bf k} 
\left( 
u_k e^{i{\bf k}\cdot {\bf r}} 
{\hat b}_{{\bf k}\uparrow} 
- 
v_k e^{-i{\bf k}\cdot {\bf r}} 
{\hat b}_{{\bf k}\downarrow} 
\right) 
\; , 
\eeq 
in terms of the anticommuting quasi-particle 
Bogoliubov operators ${\hat b}_{{\bf k}\sigma}$. 
The quasi-particle 
amplitudes $u_k$ and $v_k$, such that $u_k^2+v_k^2=1$, 
are obtained by imposing 
the minimization \cite{landau} of the thermodynamic potential   
\beq 
\Omega = \int d^3{\bf r} \; 
\langle {\hat {\cal H}}({\bf r}) - \mu \, \hat{n}({\bf r}) \rangle  \; , 
\label{ther} 
\eeq 
where $\mu$ is the chemical potential, fixed by the 
average number $N$ of fermions. 
At zero-temperature the average of quasi-particle Bogoliubov 
operators is given by 
\beq
\langle 
{\hat b}_{{\bf k}\sigma}^+ 
{\hat b}_{{\bf k}\sigma'} 
\rangle = \Theta(E_k) \, \delta_{\sigma \sigma'} 
\eeq
where $E_k$ are the quasi-particle energies and $\Theta(x)$ is the 
Heaviside step function. Neglecting the Fock terms 
$\langle {\hat \psi}_{\sigma}^+ {\hat \psi}_{\sigma'} \rangle$ 
with $\sigma \neq \sigma'$ \cite{pinilla}, 
after minimization of (\ref{ther}) 
one recovers \cite{bardeen,fetter} 
the standard BCS equation for the number of particles 
\beq 
N = 2 \sum_{\bf k} v_k^2  \; , 
\label{bcs1} 
\eeq
and the familiar BCS gap equation 
\beq
-{1\over g} = {1 \over V} \sum_{\bf k} {1\over 2 E_k} 
\label{bcsGap}
\eeq 
Here
\beq
E_k=\left[\left({\hbar^2k^2\over 2m}-\mu \right)^2 
+ \Delta^2\right]^{1/2}
\eeq
and
\beq \label{vk}
v_k^2 = {1\over 2} \left( 1 - \frac{{\hbar^2k^2\over 2m} 
- \mu }{E_k} \right) \, ,  
\eeq 
with $u_k^2=1 - v_k^2$. 
The chemical potential $\mu$ and the gap energy $\Delta$ are obtained by
solving equations (\ref{bcs1}) and (\ref{bcsGap}).
Unfortunately, in the continuum limit, due to the 
choice of a contact potential, the gap equation diverges in the 
ultraviolet. This divergence is logarithmic in two dimensions 
and linear in three dimensions. 

As discussed in \cite{marini}, quite generally in two dimensions 
the bound-state energy $\epsilon_B$ exists for any value 
of the interaction strength $g$. For the contact potential 
the bound-state equation is 
\beq 
- {1 \over g} = 
{1 \over V} \sum_{\bf k} \frac{1}
{{\hbar^2k^2\over 2m} + \epsilon_B} \, ,
\eeq 
and then subtracting this equation from 
the gap equation \cite{eagles,leggett,noziers}, 
one obtains a regularized gap equation 
\beq 
\sum_{\bf k} \left( 
\frac{1}{ {\hbar^2k^2\over 2m} + \epsilon_B} 
- {1\over 2 E_k} \right) = 0 \; . 
\label{bcs2}  
\eeq 
In the two-dimensional 
continuum limit $\sum_{\bf k} \to V/(2\pi)^2 \int d^2{\bf k}
\to V/(2\pi) \int k dk$, taking into account 
the functional dependence (\ref{vk}) of the 
amplitudes $u_k$ and $v_k$ on $\mu$ and $\Delta$, 
the Eq. (\ref{bcs2}) gives 
\beq 
\epsilon_B = \Delta 
\left( \sqrt{1+{\mu^2\over \Delta^2}}- {\mu\over \Delta} \right) \; ,  
\eeq
while the number equation (\ref{bcs1}) becomes 
\beq 
n = {N\over V} = \big({m \over 2\pi\hbar^2}\big) \Delta 
\left( {\mu\over \Delta} + \sqrt{1+{\mu^2\over \Delta^2}} \right) \; . 
\label{numb} 
\eeq
These two equations are exactly those found in the appendix B of 
the paper of Marini, Pistolesi and Strinati \cite{marini}. 
We observe that, for a 2D inter-atomic potential described 
by a 2D circularly symmetric well of radius $R_0$ and 
depth $U_0$, the bound-state energy $\epsilon_B$ is given 
by $\epsilon_B \simeq \hbar^2/(2mR_0^2) \exp{(-2\hbar^2/(mU_0R_0^2))}$ 
with $U_0 R_0^2 \to 0$ \cite{landau}.  
 
As previously stressed, several properties of ultra-cold Fermi gases 
have been investigated in the last few years by 
using the EBCS equations \cite{perali,hu}. 
Here we analyze the condensate fraction of fermionic pairs 
that is strictly related 
to the off-diagonal long-range order (ODLRO) \cite{penrose} of the system. 
As shown by Yang \cite{yang}, the BCS state 
guarantees the ODLRO of the Fermi gas, namely that, in the limit wherein
both unprimed coordinates approach an infinite distance from the primed 
coordinates, the two-body density matrix factorizes as follows: 
\beqa
\label{2bodydm}
\langle 
{\hat \psi}^+_{\uparrow}({\bf r}_1') 
{\hat \psi}^+_{\downarrow}({\bf r}_2') 
{\hat \psi}_{\downarrow}({\bf r}_1) 
{\hat \psi}_{\uparrow}({\bf r}_2)  
\rangle 
\\ \nonumber
= \langle 
{\hat \psi}^+_{\uparrow}({\bf r}_1') 
{\hat \psi}^+_{\downarrow}({\bf r}_2')  
\rangle \langle 
{\hat \psi}_{\downarrow}({\bf r}_1)  
{\hat \psi}_{\uparrow}({\bf r}_2)  
\rangle \, . 
\eeqa
The largest eigenvalue $N_0$ of the two-body density matrix (\ref{2bodydm})
gives the number of Fermi pairs in the lowest state, i.e. the condensate
number of Fermi pairs \cite{leggett,yang,campbell}. 
This number is given by 
\beq
\label{ODLRO:def}
N_0 = \int d^3{\bf r}_1 \; d^3{\bf r}_2 \; | \langle 
{\hat \psi}_{\downarrow}({\bf r}_1)  
{\hat \psi}_{\uparrow}({\bf r}_2)  
\rangle |^2 ,
\eeq 
and it is straightforward to show \cite{campbell} that 
\beq 
N_0 = \sum_{\bf k} u_k^2 v_k^2 \; . 
\eeq 
In the two-dimensional continuum limit we find 
\beq 
n_0 = {N_0\over V} = {1\over 4} \int {d^2k \over 2\pi} 
{\Delta^2 \over \left({\hbar^2k^2\over 2m} -\mu\right)^2 + \Delta^2 } \; ,    
\eeq 
from which 
\beq 
n_0 = {1\over 4} \big({m\over 2 \pi \hbar^2}\big) \Delta 
\left( {\pi\over 2} + \arctan{({\mu\over \Delta})} \right) \; . 
\label{cond}
\eeq 
Finally, by using Eq. (\ref{numb}) and Eq. (\ref{cond}) 
we obtain a remarkably simple formula for the condensed fraction 
\beq 
{n_0\over n} = {1\over 4} 
{ {\pi\over 2} + \arctan{({\mu\over \Delta})}  
\over 
{\mu\over \Delta} + \sqrt{1+{\mu^2\over \Delta^2}} }  \; ,  
\label{fraction} 
\eeq
This is the main result of the paper. 
Nicely, in Eq. (\ref{fraction}) the condensate fraction 
depends only on the parameter $x_0=\mu/\Delta$. 
In the weakly-bound BCS regime ($x_0 \gg 1$) 
the condensed fraction $n_0/n$ goes to zero, while in the 
strongly-bound BEC regime ($x_0 \ll -1$) 
the condensed fraction $n_0/n$ goes to $1/2$, i.e. all the 
$N/2$ Fermi pairs belong to the Bose-Einstein condensate. 
\par
In 2D the Fermi energy $\epsilon_F=\hbar^2k_F^2/(2m)$ 
of a non-interacting Fermi gas is given by 
$\epsilon_F=\pi \hbar^2 n/m$. It is convenient to express 
all relevant energies in terms of the Fermi energy 
$\epsilon_F$. In this way these scaled quantities depend 
only on the parameter $x_0$. In particular, we find 
\beq 
{\epsilon_B\over \epsilon_F} = 
2 { 
\sqrt{1+ x_0^2} - x_0   
\over 
\sqrt{1+ x_0^2} + x_0} \; ,  
\label{epsilonB} 
\eeq
\beq
{\Delta\over \epsilon_F} = 2 \left(\sqrt{1+ x_0^2} - x_0\right)  \; , 
\eeq
and also 
\beq
{\mu \over \epsilon_F} = 2 x_0 \left( \sqrt{1+ x_0^2} - x_0\right) \; .
\eeq 
All these quantities are parametrized by $x_0$. It is then 
quite easy to plot the scaled 
chemical potential $\mu/\epsilon_F$ and the scaled energy gap 
$\Delta/\epsilon_F$ as a function of the scaled bound-state energy 
$\epsilon_B/\epsilon_F$. 

\begin{figure}
\centerline{\epsfig{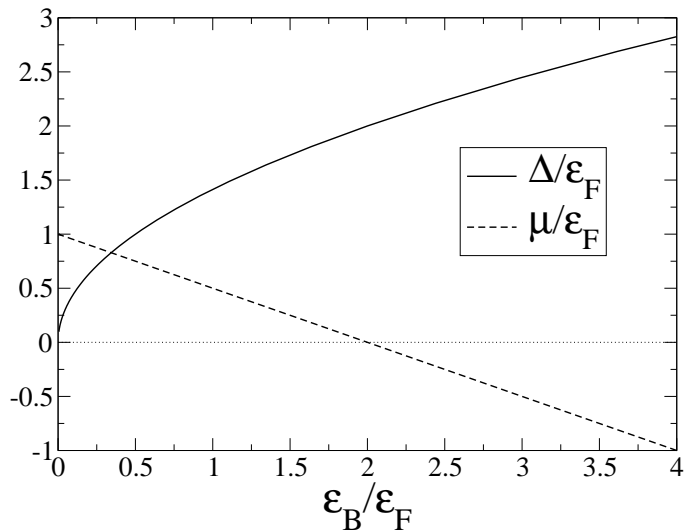}}
\small 
\caption{Energy gap $\Delta$ (solid line) 
and chemical potential $\mu$ (dashed line) 
in the uniform two-component 
dilute 2D Fermi gas as a function of scaled 
bound-state energy $\epsilon_B/\epsilon_F$. 
The horizontal dotted line simply shows the zero.} 
\end{figure} 

Fig. 1 shows that the scaled energy gap $\Delta/\epsilon_F$ 
(solid line) grows by increasing $\epsilon_B/\epsilon_F$ 
while the scaled chemical potential $\mu/\epsilon_F$ (dashed line) 
decreases. The chemical potential $\mu$ is zero 
when the bound-state energy $\epsilon_B$ is equal to $2\epsilon_F$. 
For larger values of $\epsilon_B$ the chemical potential $\mu$ 
becomes negative. Physically, we can say that 
the value $2\epsilon_F$ discriminates between the BCS regime 
of weekly-bound fermionic pairs ($0\le \epsilon_B<2\epsilon_F$ 
and $\mu>0$) and the BEC regime of strongly-bound fermionic pairs 
($\epsilon_B>2\epsilon_F$ and $\mu<0$). 
\par 
By using Eq. (\ref{fraction}) and Eq. (\ref{epsilonB}) 
we can also plot the condensate fraction $n_0/n$ 
as a function of the scaled bound-state energy 
$\epsilon_B/\epsilon_F$.

\begin{figure}
\centerline{\epsfig{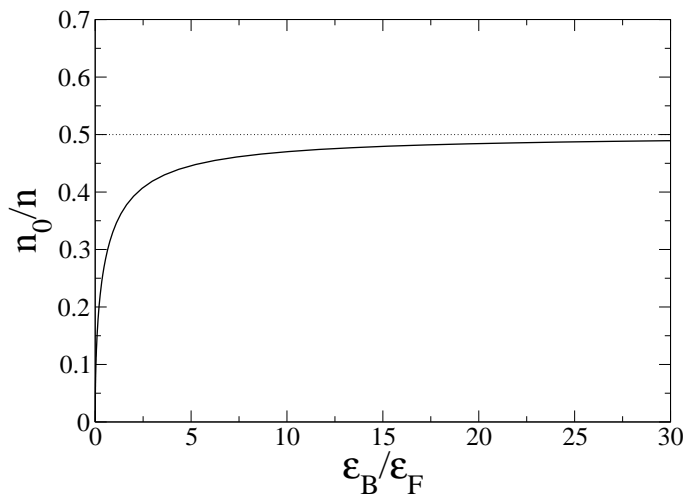}}
\small
\caption{Condensate fraction $n_0/n$ of 
Fermi pairs (solid line) in the uniform two-component 
dilute 2D Fermi gas as a function of scaled
bound-state energy $\epsilon_B/\epsilon_F$. 
The horizontal dotted line shows the asymptotic value $n_0/n=1/2$.}
\end{figure}

Fig. 2 shows the condensate fraction $n_0/n$ of Fermi pairs. 
The fraction is zero when the bound-state energy $\epsilon_B$ is zero. 
For small values of $\epsilon_B/\epsilon_F$ the condensed 
fraction has a very fast grow but then it reaches the 
asymptotic value $1/2$ very slowly. 
Note that at $\epsilon_B/\epsilon_F=2$ (where $\mu=x_0=0$) 
the condensate fraction is $\pi/8\simeq 0.39$. 
\par 
To the sake of completeness, we observe that in the 3D case 
the condensate fraction can be expressed in terms 
of the dimensionless inverse interaction parameter 
$y = (k_F a_F)^{-1}$, where $k_F$ 
is the Fermi wave vector of 3D non-interacting fermions 
and $a_F$ is the 3D s-wave scattering length of inter-atomic 
potential \cite{sala-odlro}. In this case, $y\in (-\infty,+\infty)$ 
and only for $y>0$ there is the formation 
of a dimer with bound-state energy 
$\epsilon_B \simeq \hbar^2/(2ma_F^2)$. 
The condensed fraction goes to zero for $y\to-\infty$ and to 
one-half for $y\to +\infty$ \cite{sala-odlro}.  

In conclusion, by using the mean-field extended BCS theory and 
the concept of off-diagonal long-range order, that is 
the existence of a macroscopic eigenvalue of the two-body 
density matrix, we have obtained a remarkably 
simple formula for the condensate fraction of fermionic pairs in 
a uniform 2D Fermi gas. Contrary to the 3D case, in the 2D case 
a bound state appears immediately as the two-body attraction 
is introduced. As a consequence, 
the crossover from the BCS state of weakly-bound Cooper pairs 
to the BEC of strongly-bound dimers is induced by 
the increasing of the bound-state energy. 
We have show that the condensate fraction
of Fermi pairs grows smoothly during 
the 2D BCS-BEC crossover, but only for a very 
large bound-state energy one gets a quasi complete  
condensation. It is important to stress that our predictions 
on the behavior of the condensed fraction in a 2D attractive Fermi gas 
can be surely compared with Monte Carlo calculations, 
as done in the 3D case \cite{astrakharchik}. On the other hand, 
it could be difficult to compare the theory with experiments. In fact, 
strictly 2D superfluid Fermi gases have not yet been achieved: 
a 2D configuration requires $n a_H^2\ll 1$, where $n$ is the 2D 
number density and $a_H$ is the characteristic length of a very strong 
harmonic confinement along one of the three axes \cite{sala-sadhan2}. 
In addition, in two-dimensions it could be more problematic than 
in three-dimensions to modify and trigger the bound-state energy 
\cite{scattering}. 
Surely in the next future these issues will be faced 
and probably the obstacles will be overcomed.  

The author thanks A. Parola and F. Toigo for useful discussions.

\end{document}